\documentclass[12pt]{article}
\usepackage{mathrsfs}
\newsavebox\foobox
\newlength{\foodim}

\usepackage{color}
\usepackage{amsfonts,amssymb,amsmath}
\usepackage[export]{adjustbox}
 
\usepackage{graphicx}
\usepackage[T1]{fontenc}
\usepackage[numbers,sort&compress]{natbib}
\graphicspath{ {./images/} } \textheight 9in \textwidth  6.5in
\newcounter{tempeq}
\begin{document}
\title{\textsf{Charging a Quantum Battery Mediated by Parity-Deformed Fields}}
\author{B. Mojaveri\thanks{Email: bmojaveri@azaruniv.ac.ir; bmojaveri@gmail.com (corresponding author)},
\hspace{2mm}R. Jafarzadeh Bahrbeig\thanks{Email:
r.jafarzadeh86@gmail.com} \hspace{.75mm}and\hspace{2mm}M. A. Fasihi
\thanks{Email: a.fasihi@gmail.com}\\
{\small {Department of Physics, Azarbaijan Shahid Madani University,
PO Box 51745-406, Tabriz, Iran \,}}} \maketitle
\begin{abstract}
We study the effect of parity deformation of the environmental field
modes on the wireless charging performance of a qubit-based open
quantum battery (QB) consisting of a qubit-battery and a
qubit-charger, where there is no direct interaction between the
qubits and battery is charged by the mediation of the environment.
The parity deformation introduces field nonlinearities as well as
qubit-environment intensity-dependent couplings. We analyze in
detail charging characteristics, including the charging energy,
efficiency and ergotropy in both the weak and strong coupling
regimes, and show that the memory effects of mediator environment
are critical in enhancing the charging performance. In the strong
coupling regime, parity deformation of the environment fields can
further trigger non-Markovian quantum memory of the charger-battery
system, thereby enhancing the QB charging performance based on the
non-Markovianity. Surprisingly, if the charging process is Markovian
in the absence of the parity deformation, parity deformation is able
to induce memory effects in the charger-battery dynamics and
transforms the Markovian process to the non-Markovian one. This work
highlights that proper engineering of the coupling to an environment
can introduce an extra quantum memory source to the underlying
charging process in favor of environment-mediated charging of the
battery.\\\\
{\bf Keywords:} Open quantum battery, Parity deformation,
Intensity-dependent coupling, Wireless battery charging, Ergotropy,
Non-Markovianity.
\end{abstract}
\section{Introduction}
In recent years, there has been a significant evolution in our
understanding of energy harnessing, all thanks to the remarkable
advancements achieved in the field of quantum thermodynamics. This
has led to the emergence of a new concept called quantum batteries
(QBs) \cite{Alicki}, small-scale energy storage devices. QBs are
quantum mechanical systems that can store energy temporarily in the
finite dimensional degree of freedom using the principles of quantum
mechanics \cite{Hovh, Rossini, Ando0, Liu}. Compared to
electrochemical batteries, QBs can experience fast charging times
\cite{Ficher, Ghosh, Gao0, Gyhm, Salvia} because they take advantage
of unique features like quantum coherence and quantum entanglement.
An empty QB is usually charged based on an interaction protocol
between QB itself with either an external field or a quantum system
which serves as a charger. During the interaction, it moves from a
lower energy level to a higher one and gets charged. The charging
performance of a QB is characterized by its storage energy,
efficiency, power, storage capacity as well as ergotropy (the
maximum energy extractable from the battery through cyclic unitary).
Quantum coherence and quantum entanglement have been proven to be
beneficial in charging performance of QBs \cite{Saliva00,
Allahverdyan, Nimmrichter00}. Furthermore, a powerful charging can
be generated by taking into account the collective quantum effects
of a battery with $N$ quantum cells \cite{Keck, Binderm, Binderc,
Kitaev}. For such QBs, the total power scales with $N\sqrt{N}$,
instead of (linear) $N$ \cite{Keck}. Over the past years, many
theoretical studies have been devoted on the different aspects of
QBs. These studies include proposing charging processes \cite{Farin,
Zhang, Yang00, Dou33, Cata, Fus, Fer0, Maze, Manzo, Shaghaghi00},
evaluating the charging performance \cite{Delmonte, Allahverdyan,
Safr, Dou44, Dong00, Moj33}, suggesting possible implementation
schemes \cite{Fer0, Forn, Monsel, Lv, Quach23, Bau, Devoret, Gemme,
Lai, Dou11}, and so on. A comprehensive review on quantum batteries
can be found in Ref. \cite{Coll}.

 Along with these developments, due to the inevitable interaction between a real quantum system
and its environment, the destructive effects of environmental noises
on the QB's charging and discharging processes have been
investigated \cite{Salimi, Dou00, Down}. Nowadays there is a growing
fascination with the study QBs from the perspective of open quantum
systems. When a QB is subjected into the environmental noises, its
coherence leaks into the environment, resulting in a phenomenon
known as decoherence. Unfortunately, the environment induced
decoherence often reduces the performance of QBs during charging and
discharging \cite{Farin1, Camp}. In particular, decoherence pushes
QBs towards a non-active (passive) equilibrium state during the
charging process, making it extremely challenging to extract work
from the QBs in a cyclic unitary process \cite{Barra}. Decoherence
can also affect QBs disconnected from both charger and consumption
hub, leading to self-discharge of QBs \cite{San0, Pedro, Zhu00}.
Hence, designing QBs robust to the environmental dissipations is a
key task. Recent research efforts have focused not only on studying
the effect of the environment on QBs decoherence effects
\cite{Carega00, Zakavati, Morrone, Haseli}, but also developing open
QBs that utilize quantum control techniques to enhance the stability
of their charging performance \cite{Kamin1, Dark, Squeezing}. For
instance, in \cite{Dark} the authors use dark states to achieve a
high performance, robust QB comprising an ensemble of $1/2$-spins
that are charged by another ensemble of $1/2$-spins inside a thermal
reservoir. In another work, Centrone et al \cite{Squeezing},
proposed a continuous variable QB coupled weakly to a squeezed
thermal reservoir and managed to control the steady-state charging
performance of the battery by boosting the quantum squeezing of
reservoir. In \cite{Dou11} Dou and Yang proposed a superconducting
transmon qubit-resonator QB consisting of $N$ coupled transmon
qubits as the battery and a one-dimensional transmission line
resonator as the charger. They showed that, when qubits undergo the
dephasing channel, transitions between the qubit states occur and
the decay of qubit energy into the environment is suppressed, so
that a high performance QB is achieved. They also imposed the
condition that the anharmonicity of transmon is smaller than the
transmon-resonator detuning \cite{Dou22}, and found that by tuning
transmons into a decoherence-free subspace, a resonator-qutrits QB
with a stable and efficient charging process can be realized. A
stabilization scheme based on a sequence of repeated quantum
measurements has been proposed in \cite{Sequ}. Kamin et al
\cite{Kamin1} proposed a qubit-based QB that is charged by the
mediation of a non-Markovian environment. They showed that the
memory effects are beneficial for improving charging cycle
performance. In addition to the above robust protocols, several
other control techniques have been exploited to protect the charging
cycle of QBs such as feedback control technique \cite{Mitch, Shao,
Ios}, Bang-Bang modulation of the intensity of an external
Hamiltonian \cite{Franc}, the inversely-engineered control based on
Lewis-Riesenfeld invariants \cite{Ning00}, convergent iterative
algorithm \cite{Borhan}, inhiring an auxiliary quantum system
\cite{Behzadi}, modulating the detuning between system and reservoir
\cite{Yu0}, three level stimulated Raman adiabatic passage technique
\cite{Baris}, engineering quantum environments \cite{Segal,
Morrone}, etc.

 On the other hand, according to the previous
studies on the Markovian and non-Markovian dynamics of open quantum
systems, employing the technique of reservoir engineering through
the utilization of leaky cavities with parity-deformed fields
provides novel insights into control methods for stabilizing the
entanglement and coherence of a two-qubit system against the
environmental-induced dissipations \cite{Mojt, Mojaveri1}. It has
been shown that parity deformation of environment fields has
significant capabilities in the controlling initial entanglement of
the light-matter systems. The reason behind these capabilities is
that parity deformation introduces field nonlinearities inside the
environment and also induces an intensity-dependent coupling between
the atom and environment that together control dynamics of the
system \cite{Mojaveri1}. In the last years, several works have been
devoted to explore the role of parity deformation on the dynamics of
light-matter systems \cite{Deh11, Fakhri1, Fakhri2, Deh, Scri}. In
Ref \cite{Deh}, for example, it has been proven that parity-deformed
fields can enable robust transfer of entangled states to far qubits
of a cavity-based quantum network. It also has been shown that
parity deformation can improve the power and efficiency of a quantum
heat engine \cite{Scri}. At this point it is worth to remark that
the para-particles, despite being theoretically well defined, have
not yet been experimentally confirmed. Recently, some effective
methods for simulating para-particles have been proposed in the
literature \cite{Lara3, Lara0, Lara2, Lara1}. Particularly, a
quantum system consisting of a qubit with two field modes in strong
coupling regime has been considered as a promising physical system
for simulating the parity-deformed fields \cite{Lara2}. Also,
one-dimensional arrays of coupled waveguides \cite{Lara1} is another
interesting physical system that could offer a classical simulation
of the parity-deformed fields by taking advantage of their
f-deformed realization.

 In this study, we would like to use the controlling capability of
the parity-deformed fields to boost the charging performance of open
qubit-based quantum batteries. Our attention is directed towards a
wireless-like charging process, where the charger and battery's
qubits are immersed in a zero-temperature environment modeled by a
collection of the parity-deformed field modes, and battery is
charged by the mediation of the environment without needing an
external energy source or direct connection to the charger. More
precisely, we will investigate environment-mediated charging
performance of the battery in both the Markovian (memoryless) and
non-Markovian (memory-keeping) dynamical regimes, with emphasis on
the control role of the induced field nonlinearity and the
intensity-dependent qubit-environment coupling. We will show that
quantum memory is crucial in improving charging performance of QB,
and will prove that parity deformation is able to induce memory
effects in the dynamics and to transform the Markovian process to
the non-Markovian one. The rest of the paper has been structured as
follows. The physical model, open dynamics of the charger-battery
system and an explicit expression of their evolved reduced density
matrix are given in the second section. Section 3 is devoted to
introduce and describe several figures of merit for characterizing
the performance of QBs. Section 4 discuss our results. Finally,
section 5 concludes this paper.\begin{figure} \centering
\includegraphics[keepaspectratio, width=.6\textwidth]{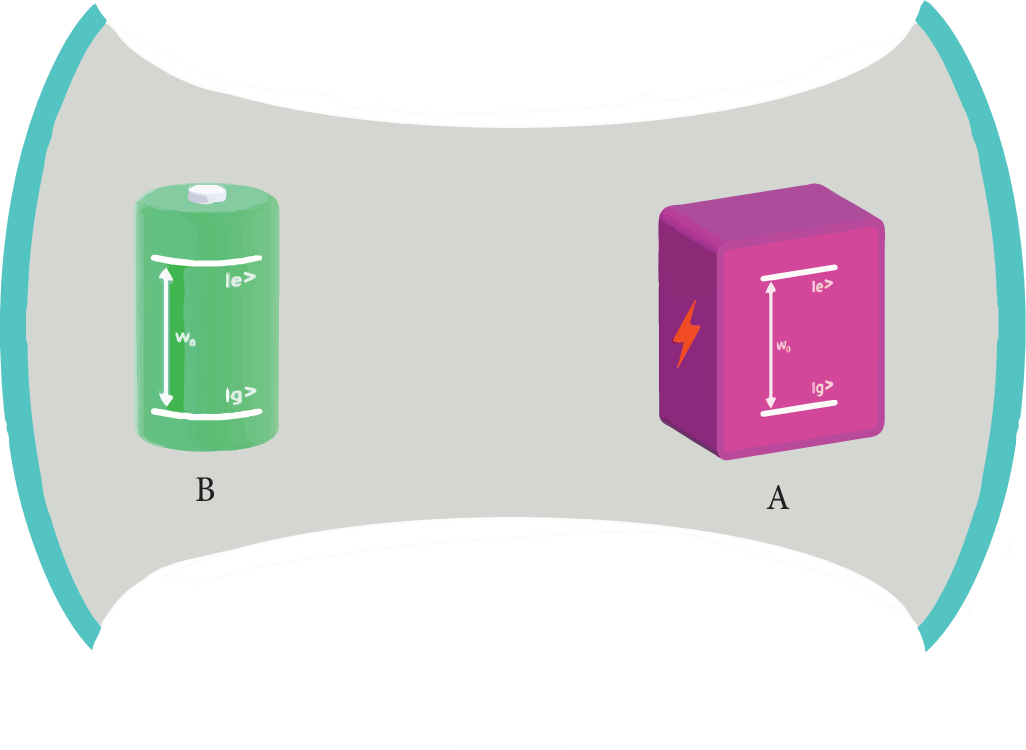}
\caption{Schematic illustration of wireless charging process of a
qubit-based QB. There is no direct interaction between the charger
and battery qubits, and battery is charged purely due to coupling
with a common dissipative para-bosonic environment.}
\end{figure}
\section{Model and Solution}
Let us consider an open charger-battery system consisting of two
qubits, the qubit A as a charger and the qubit B as a QB, both
coupled to a zero-temperature reservoir (shown in Fig. 1). Each
qubit is modeled by a two-level system with the ground state
$|g\rangle$ and excited state $|e\rangle$, but the reservoir is
modeled by a collection of the parity-deformed field modes. In our
system, there is no interaction between the qubits, and the battery
is charged by the mediation of the reservoir, which under the
rotating wave approximation is ruled by the Hamiltonian ($\hbar=1$):
\renewcommand\theequation{\arabic{tempeq}\alph{equation}}
\setcounter{equation}{-1} \addtocounter{tempeq}{1}\begin{eqnarray}
&&\hspace{-12cm}\hat{H}=\hat{H}_0+\hat{H}_{int},\label{Ho}
\end{eqnarray}
where
\renewcommand\theequation{\arabic{tempeq}\alph{equation}}
\setcounter{equation}{0} \addtocounter{tempeq}{1}\begin{eqnarray}
&&\hspace{-14mm}\hat{H}_0=\frac{\omega_0}{2}\hat{\sigma}^{(A)}_{z}+\frac{\omega_0}{2}\hat{\sigma}^{(B)}_{z}
+\sum_k\omega_k\hat{\mathbf{a}}^{\dagger}_k\hat{\mathbf{a}}_k,\\
&&\hspace{-14mm}\hat{H}_{int}=\sum_{i=A,B}\sum_k\alpha_ig_k\hat{\mathbf{a}}_k\hat{\sigma}^{(i)}_{+}+H.c.\,.
\end{eqnarray}
In the above set of equations,
$\hat{\sigma}^{(i)}_{+}=|e\rangle_i\,_i\langle g|$,
$\hat{\sigma}^{(i)}_{-}=|g\rangle_i\,_i\langle e|$ and
$\hat{\sigma}^{(i)}_{z}=\left(|e\rangle_i\,_i\langle
e|-|g\rangle_i\,_i\langle g|\right)$ $(i=A,B)$ are, respectively,
arising, lowering and inversion operators of the qubit $i$ with the
energy $\omega_0$. $\hat{\mathbf{a}}_k$ and
${\hat{\mathbf{a}}^{\dag}}_k$ are respectively the annihilation and
creation operator of the $\mathit{k}$-mode of the reservoir
para-Bose field with frequency $\omega_k$. Moreover, $\alpha_i|g_k|$
denotes the coupling strength of the qubit $i$ with mode $k$ of the
parity deformed field. The the annihilation and creation operator
$\hat{\mathbf{a}}_k$ and ${\hat{\mathbf{a}}^{\dag}}_k$ together with
the parity operator $\hat{R}_k$ constitute a deformed algebra,
namely, Wigner algebra \cite{Green0, Green1}, with the following
(anti-)commutation relations:
\renewcommand\theequation{\arabic{tempeq}\alph{equation}}
\setcounter{equation}{0} \addtocounter{tempeq}{1}
\begin{eqnarray}
&&\hspace{-14mm}\left[\hat{\mathbf{a}}_k,
\hat{\mathbf{a}}_{k^{\prime}}^{\dagger}\right]=(1+2\nu\hat{R}_k)\delta_{k,k^{\prime}},\\
&&\hspace{-14mm}\left\{\hat{R}_k,
\hat{\mathbf{a}}_{k^{\prime}}\right\}=\left\{\hat{R}_k,
\hat{\mathbf{a}}^{\dagger}_{k^{\prime}}\right\}=0,
\end{eqnarray}
where the real constant $\nu\in (-0.5,\infty)$ denotes the parity
deformation parameter. Note that, for $\nu=0$, the Wigner algebra is
reduced to the Weyl-Heisenberg algebra of the bosonic filed modes.
The Wigner algebra can be represented by the generalized Fock states
$\{\|n\rangle|_{n=0}^{\infty}\}$, as below
\renewcommand\theequation{\arabic{tempeq}\alph{equation}}
\setcounter{equation}{0} \addtocounter{tempeq}{1}
\begin{eqnarray}
&&\hspace{-14mm}\hat{\mathbf{a}}_k\left\|2n\right\rangle_k=\sqrt{2n}\left\|2n-1\right\rangle_k,\quad\quad\,\quad\quad\quad\quad\,
\hat{\mathbf{a}}_k\left\|2n+1\right\rangle_k=\sqrt{2n+2\nu+1}\left\|2n\right\rangle_k,\\
&&\hspace{-14mm}\hat{\mathbf{a}}^{\dag}_k\left\|2n\right\rangle_k=\sqrt{2n+2\nu+1}
\left\|2n+1\right\rangle_k,\quad\quad \hat{\mathbf{a}}^{\dag}_k\left\|2n+1\right\rangle_k=\sqrt{2n+2}\left\|2n+2\right\rangle_k,\\
&&\hspace{-14.5mm} \hat{R}_k\left\|n\right\rangle_k=
(-1)^n\left\|n\right\rangle_k.
\end{eqnarray}
In the photon number basis $|n\rangle$, Hamiltonian (\ref{Ho}) has
the form
\renewcommand\theequation{\arabic{tempeq}\alph{equation}}
\setcounter{equation}{-1} \addtocounter{tempeq}{1}
\begin{eqnarray}\label{NI}
&&\hspace{-1.2cm}
H=\frac{\omega_0}{2}\sum_{i=A,B}\hat{\sigma}^{(i)}_{z}+\sum_k\omega_k\hat{a}^{\dagger}_k\hat{a}_k+
\nu\sum_k\omega_k\big(1-(-1)^{\hat{n}_k}\big)+
\sum_{i=A,B}\sum_k\big(\alpha_ig_k\hat{a}_kF(\hat{n}_k)\hat{\sigma}^{(i)}_{+}+H.c.\big),\label{f-d}
\end{eqnarray}
which describes interaction of qubit-based open QB with the
reservoir fields through intensity-dependent couplings with the
intensity function
$F(\hat{n}_k)=\sqrt{\frac{\hat{n}_k+\nu\left[1-(-1)^{\hat{n}_k}\right]}{\hat{n}_k}}$
as well as in the presence of the field nonlinearities with the form
$\nu\sum_k\omega_k\big(1-(-1)^{\hat{n}_k}\big)$ \cite{Mojt,
Bashir0}. For $\nu=0$, Eq. (\ref{f-d}) is reduced to the Hamiltonian
model describing the environment-mediated open QB proposed in Ref.
\cite{Kamin1}.

 We now introduce the total excitation operator
$\hat{K}=\sum_k\hat{n}_k+\left(\hat{\sigma}_{+}^{(A)}\hat{\sigma}_{-}^{(A)}+
\hat{\sigma}_{+}^{(B)}\hat{\sigma}_{-}^{(B)}\right)$ which is a
constant of motion, i.e. $[H,\hat{K}]=0$. This allows us to
decompose Hilbert space of the entire qubit-reservoir system,
$\mathcal{H}$ spanned by the basis $|i,j\rangle\otimes|n_1,n_2,
...,n_k, ...\rangle_{\mathcal{E}}|_{n_1,n_2,...=0}^{\infty}$
$\left(i,j=e_A,g_B\right)$ into the excitation subspaces, as follows
\renewcommand\theequation{\arabic{tempeq}\alph{equation}}
\setcounter{equation}{-1} \addtocounter{tempeq}{1}
\begin{eqnarray}
&&\hspace{-14mm} \mathcal{H}=\oplus_{n=0}^{\infty} \mathcal{H}_{n},
\end{eqnarray}
where $\mathcal{H}_0$ is a null-excitation subspace spanned by the
single vector $|g,g\rangle|0\rangle_{\mathcal{E}}$ equivalent to
both the battery and charger qubits being in their ground state and
the reservoir field in the vacuum state
$|0\rangle_{\mathcal{E}}=|0,0,...\rangle_{\mathcal{E}}$. The
single-excitation subspace $\mathcal{H}_1$ has infinite dimension
because it is spanned by vectors
$\{|g_A,g_B\rangle\otimes|1_k\rangle_{\mathcal{E}}|_{k=0}^\infty,
|e_A,g_B\rangle\otimes|0\rangle_{\mathcal{E}},
|g_A,e_B\rangle\otimes|0\rangle_{\mathcal{E}}\}$ where the
excitation is either in one of the qubits or k-th mode of the
reservoir field. The remaining subspaces are:
\renewcommand\theequation{\arabic{tempeq}\alph{equation}}
\setcounter{equation}{-1} \addtocounter{tempeq}{1}{\footnotesize
\begin{eqnarray} &&\hspace{-12mm}
\tiny
\mathcal{H}_{n}=span\Big\{|g_A,g_B\rangle\otimes|n_k\rangle_{\mathcal{E}},
|e_A,g_B\rangle\otimes|n_k-1\rangle_{\mathcal{E}},|g_A,e_B\rangle\otimes|n_k-1\rangle_{\mathcal{E}},|e_A,e_B\rangle\otimes|n_k-2\rangle_{\mathcal{E}}\Big\},\,
n=2,3,...
\end{eqnarray}}
\normalsize\\ Such a decomposition enables us to restrict dynamics
of the whole qubit-reservoir system to the excitation subspaces.
Here we are interested to restrict dynamics of the entire system to
the one-excitation subspace, i.e. $\mathcal{H}_{1}$. We choose a
normalized initial state of whole qubit-reservoir system belong to
the one-excitation subspace $\mathcal{H}_{1}$ as a linear
superposition of vectors $|e_A,g_B\rangle|0\rangle_{\mathcal{E}}$
and $|g_A,e_B\rangle|0\rangle_{\mathcal{E}}$, as follows
\renewcommand\theequation{\arabic{tempeq}\alph{equation}}
\setcounter{equation}{-1} \addtocounter{tempeq}{1}\begin{equation}
|\psi(0)\rangle=\big[c_1(0) |e_{A},g_{B}\rangle +c_2(0)
|g_{A},e_{B}\rangle\big]\otimes |0\rangle_{\mathcal{E}}.
\end{equation}
where $c_1(0)$ and $c_2(0)$ are the probability amplitudes at $t=0$.
As a results, at any time $t>0$, the quantum state of the entire
qubit-reservoir system can be expressed according to the following
expression
\renewcommand\theequation{\arabic{tempeq}\alph{equation}}
\setcounter{equation}{-1}
\addtocounter{tempeq}{1}\begin{eqnarray}\label{sait}
&&\hspace{-1.2cm}|\psi(t)\rangle=
\big[c_1(t)|e_A,g_B\rangle+c_2(t)|g_A,e_B\rangle\big]\otimes|0\rangle_{\mathcal{E}}+
\sum_kc_k(t)e^{-i(-\omega_0+(2\nu+1)\omega_k)t}|g_A,g_B\rangle\otimes|1_k\rangle_{\mathcal{E}}.
\end{eqnarray}
The restriction of dynamics to the single-excitation subspace enable
us to simulate the Hamiltonian (\ref{f-d}) as a spin model
consisting of two uncoupled spins surrounded by a chain of other
spins. To show this, inspired by the idea given in \cite{Sanders0,
Somma}, we identify the qubit and reservoir operators in Hamiltonian
(\ref{f-d}) as Pauli operators, and encode the subspace
$\mathcal{H}_{1}$ to the Hilbert space of the spin model. The qubit
part is trivially mapped to spins. For the reservoir, we use a
direct one-to-one mapping and encode the bosonic basis
$|0\rangle_{\mathcal{E}}$ and $|1_k\rangle_{\mathcal{E}}$, as
follows
\renewcommand\theequation{\arabic{tempeq}\alph{equation}}
\setcounter{equation}{0} \addtocounter{tempeq}{1} \begin{eqnarray}
&&\hspace{-1.5cm}|0\rangle_{\mathcal{E}}=|0,0,...\rangle_{\mathcal{E}}\rightarrow
|\downarrow\rangle_1\otimes|\downarrow\rangle_2\otimes...,\\
&&\hspace{-1.5cm}|1_k\rangle_{\mathcal{E}}=|0,...,0_{k-1},1_k,
0_{k+1},0,..\rangle_{\mathcal{E}}\rightarrow
|\downarrow\rangle_1\otimes...\otimes|\downarrow\rangle_{k-1}\otimes|\uparrow\rangle_k\otimes
|\downarrow\rangle_{k+1}\otimes...,
\end{eqnarray}
\normalsize where $|\downarrow\rangle_k=\left(
                            \begin{array}{c}
                              0 \\
                              1 \\
                            \end{array}
                          \right)
$ and $|\uparrow\rangle_k=\left(
                            \begin{array}{c}
                              1 \\
                              0 \\
                            \end{array}
                          \right)$ are spinors of the k-th spin of the chain. Furthermore, we use the actions of the field's
operators on the subspace $\mathcal{H}_{1}$ and map them to
generators of the Pauli group,
\renewcommand\theequation{\arabic{tempeq}\alph{equation}}
\setcounter{equation}{0} \addtocounter{tempeq}{1}
\begin{eqnarray}
&&\hspace{-1.2cm}a_k\rightarrow|\downarrow\rangle_k\, _k \langle \uparrow|\equiv \hat{\sigma}^{(k)}_-,\label{Pauli1}\\
&&\hspace{-1.2cm} (-1)^{n_k}\rightarrow |\uparrow\rangle_k\, _k
\langle \uparrow|-\downarrow\rangle_k\, _k \langle
\downarrow|\equiv\hat{\sigma}^{(k)}_z.\label{Pauli2}
\end{eqnarray}
Now, according to Eqs. (\ref{Pauli1}) and (\ref{Pauli2}), the
Hamiltonian (\ref{f-d}) can be rewritten as Hamiltonian of open
quantum system consisting of two central spin $A$ and $B$ that are
surrounded by a spin environment,
\renewcommand\theequation{\arabic{tempeq}\alph{equation}}
\setcounter{equation}{-1} \addtocounter{tempeq}{1}
\begin{eqnarray}
&&\hspace{-1.2cm}
\frac{\omega_0}{2}\sum_{i=A,B}\hat{\sigma}^{(i)}_{z}+(2\nu+1)\sum_k\omega_k\hat{\sigma}^{(k)}_{+}\hat{\sigma}^{(k)}_{-}+
\sqrt{2\nu+1}\sum_{i=A,B}\sum_k\big(\alpha_ig_k\hat{\sigma}^{(k)}_{-}\hat{\sigma}^{(i)}_{+}+H.c.\big).
\end{eqnarray}
Here the field nonlinearity $\big(1-(-1)^{\hat{n}_k}\big)$ and
intensity-dependent coupling
$\big(\hat{a}_kF(\hat{n}_k)\hat{\sigma}^{(i)}_{+}+H.c.\big)$ have
been identified respectively as
$\hat{\sigma}^{(k)}_{+}\hat{\sigma}^{(k)}_{-}$ and
$\big(\hat{\sigma}^{(k)}_{-}\hat{\sigma}^{(i)}_{+}+H.c.\big)$
interactions which can be potentially implemented by the
superconducting integrated circuits \cite{Dai01, Nori123}.

 By tracing over the field modes, the reduced density operator for
the charger-battery system in the \{$|e,e\rangle$, $|e,g\rangle$ ,
$|g,e\rangle$ , $|g,g\rangle$\} basis at time $t$ is given by
\renewcommand\theequation{\arabic{tempeq}\alph{equation}}
\setcounter{equation}{-1}
\addtocounter{tempeq}{1}\begin{eqnarray}\label{ro}
  \hat{\rho}_{AB}(t)=\begin{pmatrix}
                  0 & 0 & 0 & 0 \\
                  0 & {|c_1(t)|}^2 & c_1(t){c^\ast}_2(t) & 0 \\
                  0 & {c^\ast}_1(t)c_2(t) & {|c_2(t)|}^2 & 0 \\
                  0 & 0 & 0 & 1-{|c_1(t)|}^2-{|c_2(t)|}^2
                \end{pmatrix}.
\end{eqnarray}
By solving the time-dependent Schr\"{o}dinger equation, one can
easily show that the equations of motion of the probability
amplitudes $c_1(t)$, $c_2(t)$ and $c_k(t)$ reduce to:
\renewcommand\theequation{\arabic{tempeq}\alph{equation}}
\setcounter{equation}{0} \addtocounter{tempeq}{1}
\begin{eqnarray}
&&\hspace{-14mm}i\dot{c}_1(t)=\sqrt{2\nu+1}\alpha_1\sum_kc_k(t)g_ke^{-i(-\omega_0+(2\nu+1)\omega_k)t} \label{c1}\\
&&\hspace{-14mm}i\dot{c}_2(t)=\sqrt{2\nu+1}\alpha_2\sum_kc_k(t)g_ke^{-i(-\omega_0+(2\nu+1)\omega_k)t} \label{c2}\\
&&\hspace{-14mm}i\dot{c}_k(t)=\sqrt{2\nu+1}g^{\ast}_ke^{i(-\omega_0+(2\nu+1)\omega_k)t}\Big(
\alpha_1c_1(t)+\alpha_2c_2(t)\Big). \label{ck}
\end{eqnarray}
Formal integration of Eq. (\ref{ck}) and inserting its solution into
Eqs. (\ref{c1}) and (\ref{c2}) yields
\renewcommand\theequation{\arabic{tempeq}\alph{equation}}
\setcounter{equation}{0} \addtocounter{tempeq}{1}\begin{eqnarray}
&&\hspace{-14mm}\dot{c}_1(t)=-(2\nu+1)\alpha_1\int_{0}^{t}dt^\prime f(t,t^\prime)(\alpha_1c_1(t^\prime)+\alpha_2c_2(t^\prime))\label{c1t} \\
&&\hspace{-14mm}\dot{c}_2(t)=-(2\nu+1)\alpha_2\int_{0}^{t}dt^\prime
f(t,t^\prime)(\alpha_1c_1(t^\prime)+\alpha_2c_2(t^\prime))
\label{c2t},
\end{eqnarray}
where
$f(t,t^\prime)=\sum_k{|g_k|}^2e^{-i\left[-\omega_0+(2\nu+1)\omega_k\right](t-t^\prime)}$
is the memory kernel function which in the continuum limit of the
reservoir filed modes, takes the following form
\renewcommand\theequation{\arabic{tempeq}\alph{equation}}
\setcounter{equation}{-1}
\addtocounter{tempeq}{1}\begin{eqnarray}\label{cor}
f(t,t^\prime)=\int d\omega
J(\omega)e^{-i\left[-\omega_0+(2\nu+1)\omega\right](t-t^\prime)},
\end{eqnarray}
where $J(\omega)$ denotes the spectral density of the
electromagnetic field of the reservoir. In the following, we choose
a Lorentzian spectral density \cite{Breuer0} which could be realized
by a cavity environment \cite{Lenard}. The Lorentzian spectral
density is characterized by
\renewcommand\theequation{\arabic{tempeq}\alph{equation}}
\setcounter{equation}{-1}
\addtocounter{tempeq}{1}\begin{eqnarray}\label{spec}
J(\omega)=\frac{1}{\pi}\frac{W^2\lambda}{(\omega_0-\omega-\Delta)^2+\lambda^2},
\end{eqnarray}
where $W$ is proportional to the vacuum Rabi frequency and $\Delta$
denotes the detuning of the atomic frequency $\omega_0$ and the
central frequency of the reservoir. In addition, $\lambda$ specifies
the spectral width of the coupling which is linked to the memory
time $\tau_E$ by the relation $\tau_E=\lambda^{-1}$. The relative
magnitudes of $W$ and $\lambda$, is used to distinguish the weak
coupling regime from the strong one \cite{dalton}. For $\lambda>2W$,
we have a weak coupling regime and dynamics of the charger-battery
system is Markovian. In this regime information or energy decays
irreversibly. However the strong coupling regime corresponds to
$\lambda<2W$, where the dynamics is deemed non-Markovian. In the
strong regime, the information or energy flows
back from the reservoir to the system \cite{Breuer0}.\\

 By inserting the Lorentzian spectral density Eq. (\ref{spec}) into the Eq.
(\ref{cor}), one obtains the explicit form of the correlation
function as:
\renewcommand\theequation{\arabic{tempeq}\alph{equation}}
\setcounter{equation}{-1}
\addtocounter{tempeq}{1}\begin{eqnarray}\label{cor2}
f(t-t^\prime)=\frac{W^2}{2} e^{-\lambda^{\prime} |t-t^\prime|},
\end{eqnarray}
where $\lambda^{\prime}=(2\nu+1)\overline{\lambda}-i\omega_0$ with
$\overline{\lambda}=\lambda+i(\omega_0-\Delta)$. In view of
(\ref{cor2}), performing the Laplace transformation of the Eqs.
(\ref{c1t}) and (\ref{c2t}) we have
\renewcommand\theequation{\arabic{tempeq}\alph{equation}}
\setcounter{equation}{0} \addtocounter{tempeq}{1}\begin{eqnarray}
s\mathcal{C}_1(s)=-(2\nu+1)\alpha_1\mathcal{F}(s)\big[\alpha_1\mathcal{C}_1(s)+\alpha_2\mathcal{C}_2(s)\big]+c_1(0),\label{difff1}\\
s\mathcal{C}_2(s)=-(2\nu+1)\alpha_2\mathcal{F}(s)\big[\alpha_1\mathcal{C}_1(s)+\alpha_2\mathcal{C}_2(s)\big]+c_2(0),\label{difff2}
\end{eqnarray}
where $\mathcal{C}_1(s)$ and $\mathcal{C}_2(s)$ are the Laplace
transformation of $c_1(t)$ and $c_2(t)$, respectively, and
$\mathcal{F}(s)=\frac{W^2}{2}\frac{1}{s+\lambda^{\prime}}$ denotes
the Laplace transformation of the function $f(t)$. The Eqs.
(\ref{difff1}) and (\ref{difff2}) are now algebraic and
reformulating these two equations give us the general expressions of
the Laplace transformed functions $\mathcal{C}_1(s)$ and
$\mathcal{C}_2(s)$
\renewcommand\theequation{\arabic{tempeq}\alph{equation}}
\setcounter{equation}{0} \addtocounter{tempeq}{1}\begin{eqnarray}
\mathcal{C}_1(s)=\frac{r_2\Big[r_2c_1(0)-r_1c_2(0)\Big]}{s}+r_1\Big[r_1c_1(0)+r_2c_2(0)\Big]\mathcal{P}(s),\label{css1}\\
\mathcal{C}_2(s)=-\frac{r_1\Big[r_2c_1(0)-r_1c_2(0)\Big]}{s}+r_2\Big[r_1c_1(0)+r_2c_2(0)\Big]\mathcal{P}(s),\label{css2}
\end{eqnarray}
where $r_i=\frac{\alpha_i}{\alpha_T}$ ($i=1,2$) is the relative
coupling strength satisfying $|r_1|^2+|r_2|^2=1$, and
$\mathcal{P}(s)=\frac{1}{s+(2\nu+1)\alpha_T^2\mathcal{F}(s)}$, in
which collective coupling constant is defined as
$\alpha_T=\sqrt{\alpha_1^2+\alpha_2^2}$. At this stage, for the
following discussion, we introduce the vacuum Rabi frequency
$\mathcal{R}=W\alpha_T$. In this regard, the Markovian and
non-Markovian dynamical regimes are distinguished by the
dimensionless parameter $R=\mathcal{R}/\lambda$: $R\ll 1$ means the
Markovian (memoryless) dynamics, while $R\gg 1$ corresponds to the
non-Markovian (memory-keeping) dynamics with reversible dissipation
of information \cite{Franc05}.

 Now, by inverting the Laplace transform, we arrive at the solution
for the amplitudes
\renewcommand\theequation{\arabic{tempeq}\alph{equation}}
\setcounter{equation}{0} \addtocounter{tempeq}{1}\begin{eqnarray}
&&\hspace{-3cm}c_1(t)=r_2\Big[r_2c_1(0)-r_1c_2(0)\Big]+r_1\Big[r_1c_1(0)+r_2c_2(0)\Big]p(t)\label{c(t)1},\\
&&\hspace{-3cm}c_2(t)=-r_1\Big[r_2c_1(0)-r_1c_2(0)\Big] +r_2\Big[
r_1c_1(0)+r_2c_2(0)\Big]p(t)\label{c(t)2},
\end{eqnarray}
where the survival amplitude $p(t)$ is the inverse Laplace
transformation of $\mathcal{P}(s)$ given by
\renewcommand\theequation{\arabic{tempeq}\alph{equation}}
\setcounter{equation}{-1}
\addtocounter{tempeq}{1}\begin{eqnarray}\label{hh}
p(t)=\exp(-\frac{\lambda^\prime t}{2})\bigg[\cosh(\frac{\beta
t}{2})+\frac{\lambda^\prime}{\beta}\sinh(\frac{\beta t}{2})\bigg]
\end{eqnarray}
with
\renewcommand\theequation{\arabic{tempeq}\alph{equation}}
\setcounter{equation}{-1}
\addtocounter{tempeq}{1}\begin{eqnarray}\label{q}
\beta=\bigg[\lambda^{\prime\,
2}-4(2\nu+1)\mathcal{R}^2\bigg]^{\frac{1}{2}}.
\end{eqnarray}
The survival amplitude $p(t)$, indeed, determines the dynamic
path-way of the amplitudes and thus the environment-mediated
charging process of the QB in the damped fashion. It is not
difficult to show that the survival amplitude $p(t)$ tends to zero
at sufficiently long times. In this case, the amplitudes
(\ref{c(t)1}) and (\ref{c(t)2}) reads
\renewcommand\theequation{\arabic{tempeq}\alph{equation}}
\setcounter{equation}{0} \addtocounter{tempeq}{1}\begin{eqnarray}
&&\hspace{-3cm}c_1(\infty)=r_2\Big[r_2c_1(0)-r_1c_2(0)\Big],\\
&&\hspace{-3cm}c_2(\infty)=-r_1\Big[r_2c_1(0)-r_1c_2(0)\Big],
\end{eqnarray}
which indicates that the stationary state
$\hat{\rho}_{AB}(t\rightarrow \infty)$ depends only on the initial
conditions and relative coupling strengths $r_1$ and $r_2$.
\section{Energetics of the battery}
A natural question that arises at the end of the charging process is
that how much efficiently energy has been stored in QB. To answer
this question we need to explore the average energy transferred into
the QB at the time $t$ as well as the corresponding ergotropy
\cite{Allahverdyan}, i.e., respectively, the quantities
\renewcommand\theequation{\arabic{tempeq}\alph{equation}}
\setcounter{equation}{0} \addtocounter{tempeq}{1}\begin{eqnarray}
&&\hspace{-3cm}\Delta
E_B=\texttt{Tr}\{\rho_B(t)H_B\}-\texttt{Tr}\{\rho_B(0)
H_B\},\label{energy}\\
&&\hspace{-3cm}\mathcal{W}=\texttt{Tr}\{\rho_B(t)H_B\}-\texttt{min}_{U_B}\,\texttt{Tr}\{U_B\rho_B(t)U_B^{\dagger}H_B\},\label{ergotropy}
\end{eqnarray}
where minimization in Eq. (\ref{ergotropy}) is taken over all the
unitary transformations $U$ acting locally on the
$\rho_B(t)=\mathrm{tr}_{A}\big(\rho_{AB}(t)\big) $. By considering
the spectral decomposition of the Hamiltonian $H_B$ of the battery:
$H_B|\varepsilon_{i}\rangle=\varepsilon_{i}|\varepsilon_{i}\rangle$
with the ordered eigenvalues as
$\varepsilon_{i}\leq\varepsilon_{i+1}$, and the instantaneous
spectral decomposition $r_{i+1}\leq r_{i}$ of the instantaneous
battery state $\rho_B(t)$, associated to eigenvectors
$|r_{i}(t)\rangle$, the ergotropy can be re-expressed as
\renewcommand\theequation{\arabic{tempeq}\alph{equation}}
\setcounter{equation}{-1} \addtocounter{tempeq}{1}\begin{eqnarray}
\mathcal{W}=\sum_{i,j} r_j(t) \varepsilon_i(|\langle
r_j(t)|\varepsilon_i\rangle|^2-\delta_{ij}).
\end{eqnarray}
On the other hand, the efficiency, denoted as $\eta$, is another
figure of merit that indicates the proportion of energy that can be
extracted as the ergotropy compared to the total energy $\Delta E_B$
stored during the charging process. This metric is useful in
evaluating how effectively usable energy is being harnessed from the
total input energy during the charging process of a system. It can
be expressed as follows
\renewcommand\theequation{\arabic{tempeq}\alph{equation}}
\setcounter{equation}{-1} \addtocounter{tempeq}{1}\begin{eqnarray}
\eta=\frac{\mathcal{W}}{\Delta E_{B}}.
\end{eqnarray}
\begin{figure} \centering
\includegraphics[keepaspectratio, width=1\textwidth]{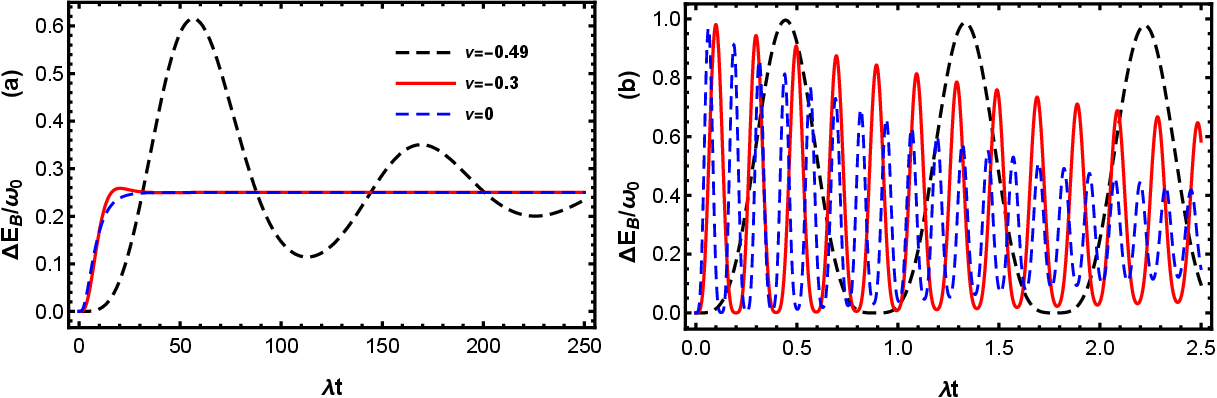}
\caption{Dynamics of the stored energy $\Delta E_B(t)$ for the
different values of $\nu$ by setting
$\Delta=\frac{2\nu}{2\nu+1}\omega_0$ and $r_1=r_2=1/\sqrt{2}$. The
panels (a) displays the Markovian dynamics with $R=0.4$, while the
panels (b) displays the non-Markovian dynamic with
$R=50$.}\label{ener}
\end{figure}
\section{Results and discussion}
In this section, we will analyze the Markovian and non-Markovian
charging performance of the introduced open QB. In particular, we
investigate the role of parity deformation of mediator environment
as well as the coupling strength of qubits with the mediator on the
dynamical behavior of performance indicators including stored
energy, ergotropy and efficiency. In what follows, we consider an
initial condition in which the battery is empty and the charger has
the maximum energy, i.e. $c_1(0)=1$, $c_2(0)=0$.\begin{figure}
\centering
\includegraphics[keepaspectratio, width=1\textwidth]{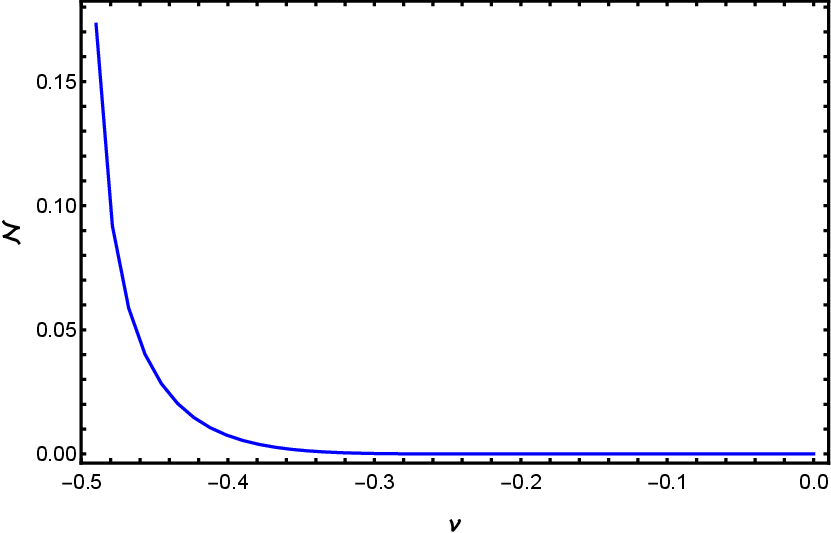}
\caption{Non-Markovianity as a function of $\nu$ for $R=0.4$. Other
parameters are taken as in Fig. 2.}\label{nonm}
\end{figure}

 In Fig. 2, we plot the stored energy $\Delta E_B$ as a function of
the dimensionless quantity $\lambda t$ for different values of the
deformation parameter $\nu$ by choosing
$\Delta=\frac{2\nu}{2\nu+1}\omega_0$ and $r_1=r_2=1/\sqrt{2}$. The
rationale for choosing $r_1=r_2=1/\sqrt{2}$ is that, as we will show
later, the QB's performance is optimized only for these specific
parameter values. On the other hand, from Eqs. (\ref{hh}) and
(\ref{q}) one can check that by setting
$\Delta=\frac{2\nu}{2\nu+1}\omega_0$ charging process of the QB is
acutely independent of the qubit's transition frequency $\omega_0$
and thus our following results will be valid in different regimes of
parameter. Furthermore, since in Ref. \cite{Kamin1} the charger and
battery's qubits are symmetrically coupled to the environment
($r_1=r_2=1/\sqrt{2}$) and also the qubit's transition frequency is
in resonant with the central frequency of environmental modes
($\Delta=0$), its results can be now recovered in the limit $\nu=0$.
In panel (a), the battery is charged by the mediation of a Markovian
environment $(R=0.4)$, while in panel (b), it is charged by the
mediation of a non-Markovian environment $(R=50)$. What is shown is
that parity deformation of the mediator fields has positive impact
in controlling the stored energy of battery in both the Markovian
and non-Markovian charging regimes. In the Markovian scenario, as
illustrated in panel (a), depending on whether $\nu=0$ or
$\nu\neq0$, different behaviors of $\Delta E_B$ emerge: in the
$\nu=0$ case, $\Delta E_B$ shows a generally monotonic increasing
behavior, and $\nu<0$ shows a non-monotonous behavior, similar to
what can be seen in panel (b). The non-monotonic behavior as a
manifestation of non-Markovian dynamics indicates that the parity
deformation introduces an enhancement of memory effects, and causes
a crossover from a Markovian to a non-Markovian in the dynamics.
Mathematically, the origin of this crossover should be traced back
to the survival amplitude $p(t)$ given in (\ref{hh}) because this
function, which generally determines all possible trajectories of
the system dynamics in the damped fashion, depends on the
deformation parameter $\nu$ by the Eq. (\ref{q}), and enhancement of
memory effects due to the parity deformation of the mediator fields
is potentially possible. We will investigate this possibility below
in the strong coupling regime.
\begin{figure}
\centering
\includegraphics[keepaspectratio, width=1\textwidth]{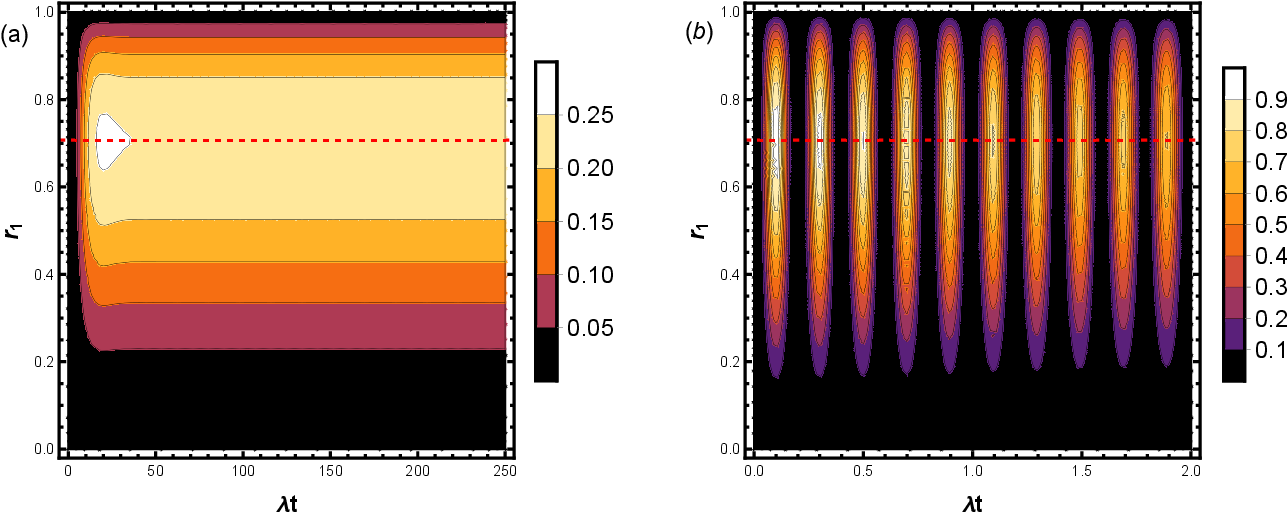}
\caption{Dynamics of the stored energy $\Delta E_B(t)$ as function
of the relative coupling strength $r_1$ for (a) Markovian dynamics
with $R=0.4$ and (b) non-Markovian dynamics with $R=50$ by setting
$\nu=-0.3$. The other parameters are taken as in Fig. 2.}\label{erg}
\end{figure}

 However, in the non-Markovian scenario, as illustrated in panel (b),
$\Delta E_B$ decays oscillatory to its steady-state value (with the
chosen relative coupling strengths $r_1=r_2=1/\sqrt{2}$, it is
$0.25$) for any $\nu$. We observe that by regularly increasing the
negative values of the parity deformation parameter, the rate of
non-Markovian oscillatory decays decreases and a large amount of the
released initial energy from the charger is transformed to the
battery, despite the lack of a direct interaction between charger
and battery's qubits. The dynamical behavior of $\Delta E_B$ in Fig.
2 can be understood from perspective of the non-Markovian nature of
the mediator reservoir, because in a environment-mediated charging
process it has been shown that transferring a large amount of energy
from the charger to the battery requires an environment with a high
degree of non-Markovianity \cite{Kamin1}. Accordingly, when the
mediator fields are deformed, increase in the negative values of the
deformation parameter seems increases the non-Markovianity of the
mediator and thus improves the memory effects of the mediator in
favor of wireless charging of the battery. To confirm this
prediction, we now investigate the effect of parity deformation on
the non-Markovianity of the mediator environment. To quantify the
non-Markovianity we employ the Breuer-Laine-Piilo (BLP) measure
\cite{BLP}, based on the distinguishability of two different initial
quantum states of an open system. The distinguishability of a given
pair of system states $\rho_1$ and $\rho_2$ is quantified by the
trace distance $D(\rho_1,\rho_2)=Tr|\rho_1-\rho_1|/2$ with
$|X|=\sqrt{X^{\dagger}X}$, and non-Markovian dynamics, i.e., the
flow of information from the environment to the system, is
represented by its time derivative
\renewcommand\theequation{\arabic{tempeq}\alph{equation}}
\setcounter{equation}{-1} \addtocounter{tempeq}{1}\begin{eqnarray}
\sigma(\rho_{1,2}(0);t)=\frac{d}{dt}D(\rho_1(t),\rho_2(t))>0.
\end{eqnarray}
The BLP measure is defined based on growth of the trace distance
over each time interval $(t_i, t^{\prime}_i)$ as follows
\renewcommand\theequation{\arabic{tempeq}\alph{equation}}
\setcounter{equation}{-1}
\addtocounter{tempeq}{1}\begin{eqnarray}\label{BLP}
&&\hspace{-3cm}\mathcal{N}=\mathrm{max}_{\rho_{1}(0),\rho_2(0)}\sum_i\big[D(\rho_1(t^{\prime}_i),\rho_2(t^{\prime}_i))-D(\rho_1(t_i),\rho_2(t_i))\big].
\end{eqnarray}
In order to calculate $\mathcal{N}$ one should determine for any
pair of initial states $\rho_{1,2}(0)$ the positive change of trace
distance in the interval $(t_i, t^{\prime}_i)$, then sum up the
contributions of all intervals, and finally maximize over the
possible initial pairs.

 In Fig. (3) we plot the $\mathcal{N}$ as function of deformation parameter $\nu$
for an optimal pair of initial state associated to the
charger-battery system. Here we choose $R=0.4$ to ensure the memory
effects of the mediator are not prominent. As shown in the figure,
the system exhibits Markovian dynamics until the negative values of
$\nu$ are small and below a certain threshold (e.g., $\nu=-0.3$ in
the figure). However, when $\nu$ exceeds this threshold the
Markovian dynamics of the charger-battery system changes to
non-Markovian (i.e., $\mathcal{N}>0$) signifying enhancement of
memory effects of the environmental noises. In this situation,
non-Markovianity increases asymptotically by regular increasing
negative values of the parity deformation parameter. Therefore,
parity deformation is able to enhance the memory effects of the
mediator environment on the charging process.\begin{figure}
\centering
\includegraphics[keepaspectratio, width=1\textwidth]{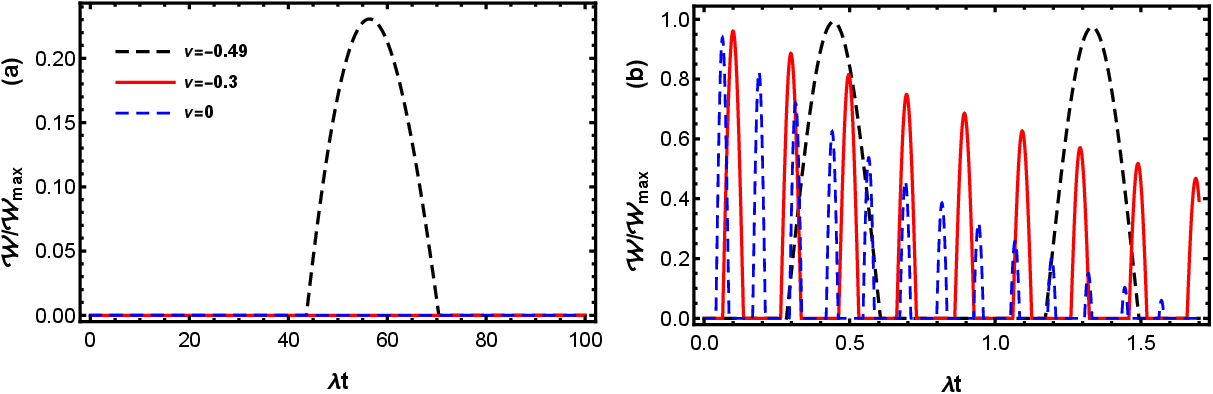}
\caption{Dynamics of ergotropy $\mathcal{W}$ for the different
values of $\nu$. The panels (a) displays the Markovian dynamics with
$R=0.4$, while the panels (b) displays the non-Markovian dynamic
with $R=50$. The parameters are taken as in Fig. 2.}\label{erg}
\end{figure}

 Now, we examine the influence of the relative coupling
strength $r_1$ on the charging performance of QB. In Fig. 4 we plot
the stored energy $\Delta E_B$ as a function of $\lambda t$ and
$r_1$ in the Markovian  (panel (a)) and non-Markovian (panel (b))
charging processes. This figure clearly shows that, for $r_1=0$ and
1, no energy stored in the QB as these correspond to cases where
only one qubit interacts with the mediator environment and there is
no correlation between qubits due to the environment. In both the
Markovian and non-Markovian charging processes, the maximum value of
stored energy is achieved at $r_1\approx0.71=\frac{1}{\sqrt{2}}$,
i.e., when the charger and battery's qubits are symmetrically
coupled to the environment. Comparing panel (a) with (b) reveals
fundamental differences of the Markovian and non-Markovian charging
processes. In the Markovian charging process, $\Delta E_B$ always
increases monotonically in time and reaches a steady-state value,
while in the non-Markovian process it experiences an oscillatory
behavior due to the memory effects of the environment. We find that
the memory effects improve energy storage and significantly increase
the charging speed. In the Markovian charging process, the battery
can be charged up to a $30\%$ of its capacity, and the optimal time
for this charge is about $\lambda t\approx 15$. However, in the
non-Markovian process the battery can be full charged, and optimal
charging time is about $\lambda t\approx 0.1$.\begin{figure}
\centering
\includegraphics[keepaspectratio, width=1\textwidth]{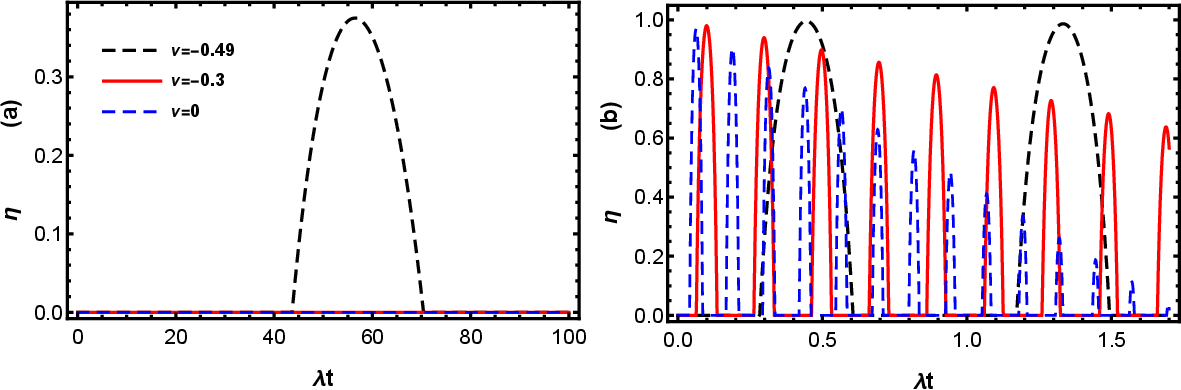}
\caption{Dynamics of efficiency $\eta$ for the different values of
$\nu$. The panel (a) displays the non-Markovian dynamics with
$R=0.4$, while the panel (b) displays the Markovian dynamic with
$R=50$. The parameters are taken as in Fig. 2.}\label{effi}
\end{figure}

 In the following, we explore the influence of parity deformation on
the dynamics of ergotropy. In Fig. 5, we plot
$\mathcal{W}/\mathcal{W}_{max}$ as a function of $\lambda t$ for the
different values of $\nu$ in the Markovian (panel (a)) and
non-Markovian (panel (b)) regimes, where the parameters are the same
as those of Fig. 2. According to the drawn curves, the effect of
parity deformation of mediator fields on the ergotropy is
constructive in both Markovian and non-Markovian dynamical regimes.
Fig. 5(a) shows that, in the Markovian regime, extracting work from
the battery is possible only when it is charged with mediation of
the parity deformed fields. In this case, whenever the deformation
parameter $\nu$ is adjusted near $-0.5$, maximal work will be
extracted. Our numerical results in Fig. 5(a) tell that, the
positive role of the parity deformation of mediator fields in the
work extraction process is prominent when we not take into account
the memory effects of the mediator. We find that, in the
non-Markovian regime regularly increasing the negative values of the
nonlinearity strength, not only enhances the ergotropy, but also
allows extracting the work over long times. Accordingly, by
benefiting from both the parity deformation as well as memory
effects of the mediator environment, a strong robust wireless-like
charging process can be established, where the extractable work
approaches to its maximum value.

 Finally, we examine the effect of parity deformation of mediator fields on the dynamics of the battery
efficiency. The results for Markovian and non-Markovian charging
processes are illustrated in Fig. 6(a) and 6(b), respectively. Here
we consider the same parameter values as in Fig. 2. Comparing Fig. 6
with 5, we can see that both efficiency and ergotropy are positively
affected by the parity deformation of mediator fields. However the
efficiency is influenced more than the ergotropy; the amount of
increment in efficiency is more than the ergotropy in both Markovian
and non-Markovian charging processes.
\section{Outlook and summary}
In summary, we proposed a novel mechanism for environment-mediated
charging process of an open QB whose performance can be well
controlled by parity deformation of the environment fields. The
system consists of a qubit-battery and a qubit-charger, where the
battery is charged by mediation of the environment fields. The
parity deformation induces a field nonlinearity of the form
$\nu\omega_k(1-(-1)^{\hat{n}_k})$ in the environment and also
introduces intensity-dependent couplings between the environment and
qubits with the intensity function
$F(\hat{n}_k)=\sqrt{\frac{\hat{n}_k+\nu\left[1-(-1)^{\hat{n}_k}\right]}{\hat{n}_k}}$.
We studied in detail the influence of the intensity-dependent
coupling and field nonlinearities induced by the parity deformation
of the mediator field modes on the charging performance, by
examining the stored energy, ergotropy and efficiency of the QB. Our
numerical findings showed that the battery has the capability to
charge wirelessly in both Markovian (weak coupling) and
non-Markovian (strong coupling) regimes. Nevertheless some
fundamental differences between Markovian and non-Markovian charging
processes were revealed. For example, the maximal amount of stored
energy in the non-Markovian charging process is more than that of
the Markovian case. Beyond that, in the Markovian regime extracting
work from the battery is possible only when the deformation
parameter $\nu$ is adjusted near its minimum value $-0.5$. To
understand the origin of these differences we studied the
non-Markovianity of underlying charging process and found that the
memory effects are crucial in enhancing the charging performance. By
increasing the negative values of the parity deformation parameter
$\nu$, the memory effects of the mediator for the non-Markovian
charging process also increases and thus the stored energy,
ergotropy and efficiency of the QB are improved. Moreover, when the
deformed parameter is set in near its minimum value, parity
deformation introduces a quantum memory source for the Markovian
charging process and transforms the Markovian process to the
non-Markovian one, thereby improving its charging performance.

 Our results showed that parity deformation is a powerful
and effective tool to control charging performance of open quantum
batteries in both the Markovian and non-Markovian dynamics. It is
worth to notice that the proposed environment-mediated charging
process has the advantage to reflect in a clear way the effects of
parity deformation on the dynamics of a QB and, at the same time, is
simple enough to make it feasible in current experimental
technologies, for instance in cavity- or circuit QED \cite{Lara3, Lara2, Strauch23} or in arrays of coupled waveguides \cite{Lara0, Lara1}.\\\\
\textbf{\large{Acknowledgment}}\\
This Work has been financially supported by Azarbaijan Shahid Madani
University under the grant number $1402/631$.\\\\
\textbf{\large{Data availability}}\\ The datasets used and analysed
during the current study available from the corresponding author on
reasonable request.

\end{document}